**Title**

Ubiquitous Knowledge empowers the Smart Factory: the impacts of a Service-oriented Digital Twin on enterprises' performance

**Abstract**

While the Industry 4.0 is idolizing the potential of an artificial intelligence embedded into "things", it is neglecting the role of the human component, which is still indispensable in different manufacturing activities, such as a machine setup or maintenance operations. The present research study first proposes an Industrial Internet pyramid as emergent human-centric manufacturing paradigm within Industry 4.0 in which central is the role of a Ubiquitous Knowledge about the manufacturing system intuitively accessed and used by the manufacturing employees. Second, the prototype of a Service-oriented Digital Twin, which leverage on a flexible ontology-oriented knowledge structure and on augmented reality combined to a vocal interaction system for an intuitive knowledge retrieval and fruition, has been designed and developed to deliver this manufacturing knowledge. Two test-beds, complimentary for the problems in practice (the former on the maintenance-production interface in a large enterprise, the latter majorly focused in production and setups in a small and medium enterprise), show the significant benefits in terms of time, costs and process quality, thus validating the approach proposed. This research shows that a human-centric and knowledge-driven approach can drive the performance of Industry 4.0 initiatives and lead a Smart Factory towards its full potential.

## 1. Introduction

The current excitement for the 4[th] Industrial Revolution can be deemed to be the idolization of an intelligence embedded into physical manufacturing systems. The concept of a ubiquitous factory (Yoon et al., 2012) characterized by a computing technology that "recedes into the background of our lives" (Weiser, 1991) is unquestionably on the forefront of technological developments and advancements in the Industry 4.0 era. Information on the manufacturing system's state has no more a single central data source. Sensors, machines, equipment, products, etc., are equipped with embedded local intelligence – which makes them smart objects – and are invisibly (over the cloud) interwoven in order to cooperate and negotiate with each other, thus becoming capable to reconfigure automatically themselves (through actuators) for flexible production of multiple types of products (Wang et al., 2016). Thanks to their increasing autonomy, the vision of the Digital Twin (or cyber twin) of a manufacturing system has become real. Firstly introduced by Grieves in 2003, the concept of Digital Twin has been adopted in the aeronautic field as a system that "mirrors the life of its flying twin" (Shafto et al., 2010). Considered as the "next wave in modelling, simulation and optimization technology" (Rosen et al., 2015), the "twinning" process – i.e. the development of a digital copy of a real system – is being recently proposed at different stages of a product life cycle (Schleich et al., 2017; Tao et al., 2018) as a linked collection of relevant digital artefacts via several simulation models (Boschert & Rosen, 2016). In this sense, the Digital Twin refers to a comprehensive physical and functional description of a component, product or system (Tuegel et al., 2011), which includes a wealth of information that could be useful in the current and subsequent life cycle phases (Cerrone et al., 2014; Söderberg et al., 2017). While, on one hand, the smart objects open opportunities for the manufacturing systems' self-regulation and real-time simulation-based optimization of products and production processes (Stark et al., 2017), Digital Twins are also emerging as the key enablers of prognostics and health management (Lee et al., 2015) and of smart maintenance, repair and overhaul activities (Qi & Tao, 2018). Scientific literature and industrial applications show how Digital Twins integrate, reuse, create and manage knowledge derived from sensed data processed via intelligent algorithms. However, streamlined knowledge access and fruition is an aspect generally overlooked in the design phase of a Digital Twin as human operators are not integral part of the 4.0 revolution. Despite automation and self-regulation by self-conscious systems seem to take over flesh-and-blood human workers are still needed, especially in operating and maintaining the industrial equipment. Also, the human workers are still more flexible than robots and they can adapt to the workflow changes much quicker, especially in mass production assembly lines. As stated in April 2018 by Elon Musk, founder & CEO of Tesla, on his Twitter: "humans are underrated - excessive automation at Tesla was a mistake" (elonmusk, 2018), commenting the delay in production of company's next car Model 3. The role of the Industrial Internet of Things should not be fostering a pure machine-to-machine manufacturing environment, but rather to enhance and optimize human work by creating new valuable data and information streams. The Industrial Internet Consortium (2016) defined the Industrial Internet (one of the different terms used interchangeably to Industry 4.0) as "an internet of things, machines, computers and people enabling intelligent industrial operations using advanced data analytics for transformational business outcomes". However, most companies often limit their efforts to the adoption of Cloud Computing and the Internet of Things (Moeuf et al., 2018) and disregard the human component and the impacts of this negligence on the production and business performance. The Industrial Internet of Things is driving the upgrade of the business models of manufacturing companies within several industries (Arnold et al., 2016). Industrial heavyweights and large enterprises, such as General Electric, have embarked on new "Industrial Internet" initiatives to realign and embed their information technology capabilities into physical equipment to offer value added services and obtain economic benefits (Agarwal & Brem, 2015). As part of their research and development initiatives, large enterprises can afford to promote and implement to a certain extent Industry 4.0 driven initiatives in different areas of their organization. Small and medium enterprises are instead hesitant in fostering and implementing such initiatives mainly because of their limited budget and their need to have a quick return on investment. A literature analysis by Moeuf et al. (2018) showed how reported Industry 4.0 projects in small and medium enterprises remain cost-driven initiatives and there is still no evidence of real business model transformation at this time. Indeed, technological-based change cannot yield long-lasting competitive advantage and sustained benefits because it can be easily imitated by other enterprises, thus feeding an "enthusiastic skepticism" on the practical benefits of Industry 4.0. The lack of a fully coupled human-machine interaction and of a cyber-human convergence generates an inadequate utilization of available information, which erodes the enterprises' production and business performance in a time when:

i. each decision or operation is directly affected by the information quality, the process uncertainties and the speed at which information reaches the end-users (Lee et al., 2014);

ii. the industrial workers need to develop new competencies that enable them to act alongside complex technological and manufacturing systems (Ras et al., 2017).

Hence, the technocratic view of the Smart Factory as a "factory-of-things" (Zuehlke, 2010) needs to advance to the next step where a human-centric manufacturing paradigm emerges within the Industry 4.0.

**Research question and study aims**

The literature review and current industrial arena has spawned then the following question: if the human operators are a non-negligible component in a manufacturing environment, how can they complement (and be integrated with) the Cyber-Physical Production System (CPPS), in a human-in-the-loop perspective, to achieve a better production and business performance? The Digital Twin appears to be the most suitable source of knowledge within the Smart Factory but the way this ubiquitous knowledge can be accessed and harnessed has not been properly investigated in the literature. Given the research question above, the study aim is to extend the current interpretation of a technology-driven Industry 4.0 towards a more human-centric paradigm where manufacturing employees (both enterprise managers and shop-floor operators) empowered with a ubiquitous knowledge about the manufacturing system are integral part of the Smart Factory and pave the way for production and business performance improvement opportunities.

The present research work proposes the concept of an Industrial Internet pyramid (described in Section 2). It is intended as an emergent human-centric manufacturing paradigm within Industry 4.0 where a Service-oriented Digital Twin acts as a link between the CPPS and the manufacturing employees. The aim is to deliver ubiquitously valuable knowledge created/managed by the Digital Twin straight to the manufacturing employees and, therefore, to achieve full information symmetry within and among all the elements of a smart factory (the cyber-physical production system downstream, the manufacturing employees upstream). The methodological and technological frameworks of the Service-oriented Digital Twin are presented in Section 3, based on which an easy to use and flexible application prototype has been implemented. Two test-beds, complimentary for the problems in practice (the former on the maintenance-production interface in a large enterprise, the latter majorly focused in production and setups in a small and medium enterprise), show the significant benefits in terms of time, costs and process quality as presented in Section 4, thus validating the approach proposed.

Being aware of the current state of the art, the following elements can be considered as the main factors proving the innovative character of the paper:

i.   a human-centric manufacturing paradigm emergent within the Industry 4.0 described in the concept of the Industrial Internet pyramid. This framework remarks the main components of an Industry 4.0 architecture but focuses on the need to deliver ubiquitously valuable knowledge created/managed by the Digital Twin straight to the manufacturing employees in order to fully exploit the potential of Digital Twin itself and Industry 4.0 initiatives;

ii.  the design, development and implementation of the prototype of a Service-oriented Digital Twin. In the Everything-as-a-Service era, a process of servitization of the Digital Twin has timidly started but the discussion is at its infancy and no technological solutions or human-centric applications have been proposed yet (Qi et al., 2018). This paper aims at making a step forward, leading to an incremental improvement in the vision of the Smart Factory, by exploring how the wealth of data and knowledge produced by the smart technological layer of a manufacturing system can be used in practice to achieve a better production and business performance. Through the use of apps and services provided by the Service-oriented Digital Twin based on a flexible knowledge structure and on augmented reality technologies combined to a vocal interaction system (a personal assistant) for an intuitive knowledge retrieval and fruition, manufacturing employees can have ubiquitous, quick and intuitive access to the manufacturing knowledge.

## 2. The Industrial Internet pyramid

Technology adoption and artificial intelligence paradigms have been the driving force of the Industry 4.0 wave, in which autonomous smart objects gave birth to CPPSs, well described in different research works (Lee et al., 2015). Today's industrial control systems are typically structured in a hierarchical manner according to the automation pyramid specified in IEC 62264, the international standard for the integration of enterprise and control systems. Whereas Level 0 of this pyramid is dedicated to the process to be controlled, which comprises all the physical objects (e.g. machines, products), Level 1 includes the intelligence embedded into these objects, thus making them "smart". Smart objects, equipped with a variable level of intelligence, are able to control and self-regulate themselves through sensors and actuators. At Level 2, the smart objects are networked with each other via communication technologies to form distributed control systems (SCADA) that monitor, collect, exchange and analyze data coming from subordinated PLCs. Level 2 can be considered then the level of the Industrial Internet of Things. However, networked devices may not be integrated because they operate on different temporal or spatial scales or because they exhibit multiple and distinct behaviors. CPPS are created only when the autonomous physical components get strongly intertwined and collaborate with each other in an intensive connection with the surrounding context and in situation dependent ways on and across all levels of production. At Level 3, integration is the key enabler of the CPPS. Several systems and interfaces can be built upon the CPPS to enable production management capabilities (with Manufacturing Execution Systems, MES) or enterprise management capabilities (with Enterprise Resource Planning systems, ERP), thus reaching the apex of the automation pyramid.

While the traditional automation pyramid describes the control systems integration within an industrial environment, it also overlooks the role of the manufacturing employees that need to be in the loop with the information streams in a ubiquitous manner. In this research work, the automation pyramid is extended to make pivotal the role of knowledge-empowered manufacturing employees in the Smart Factory. The ubiquitous access and intuitive fruition by the manufacturing employees of the knowledge about the manufacturing system's history, its real-time state and its predicted future behavior based on the sensed and processed data from the CPPS is indeed the missing (or less investigated) element of the current interpretation of a Smart Factory, today merely considered as an automated factory. The Industrial Internet pyramid, depicted in Figure 1, is grounded on the previous levels and extends the "information symmetry" of the CPPSs to the manufacturing employees. Data are collected, synthetized into information and made available to the manufacturing employees that can use this knowledge to operate back onto the system, to maintain the physical objects, to adjust the workflow and plans or to manage properly the whole manufacturing system. Such knowledge has to be "ubiquitous" as manufacturing employees require accessing it intuitively and quickly everywhere and at any time. To retrieve the information from the CPPS, process and deliver it to the end-users' fingertips (and vice versa) to let manufacturing employees interoperate as a unique body with the CPPS, a Service-oriented Digital Twin is seen as the most promising approach. We define a Service-oriented Digital Twin as the component that offers an advanced and intuitive knowledge fruition through ubiquitously accessible apps and services. Servitization (Chen, 2015) within smart manufacturing systems is gaining more and more attention because of the advantages of interoperability and platform independence (Tao & Qi, 2017). Yet, recent works only investigate the use of machine-to-machine services for mutual communication and interoperation and limit the discussion to the Level 3. The Service-oriented Digital Twin is instead made up of models, apps and services enabling the CPPS and the people using multiple data, resources and distributed computing power for ubiquitous knowledge access and advanced fruition. The vision of a Service-oriented Digital Twin represents Level 4 in the Industrial Internet pyramid, enabled by the capability to deliver a ubiquitous knowledge to the manufacturing employees. The implementation of a Service-oriented Digital Twin is on the edge and innovative (Qi et al., 2018), but it is not the Smart Factory's keystone though. It does not have any effect on the production and business performance if the manufacturing employees do not use it extensively. The manufacturing employees become then the cornerstone of the Industrial Internet pyramid. They are required to use every day on the job the new groundbreaking tools that technological advancement makes available but the switching costs are hard to overcome. Employees are usually reluctant to change especially if they do not see a practical impact on their everyday job. For this reason, the present work aims at showing how the wealth of data and knowledge produced by the smart technological layer of a manufacturing system can be used in practice to achieve a better production and business performance through a Service-oriented Digital Twin. The manufacturing employees are therefore the last component that enables the Smart Factory (Level 5 of the Industrial Internet pyramid). This pyramid intends to highlight that a human-centric manufacturing paradigm is emerging within the Industry 4.0 framework, characterized by a pervasive information symmetry on and across all the levels of the Industrial Internet pyramid is needed to achieve a significant production and business performance improvement.

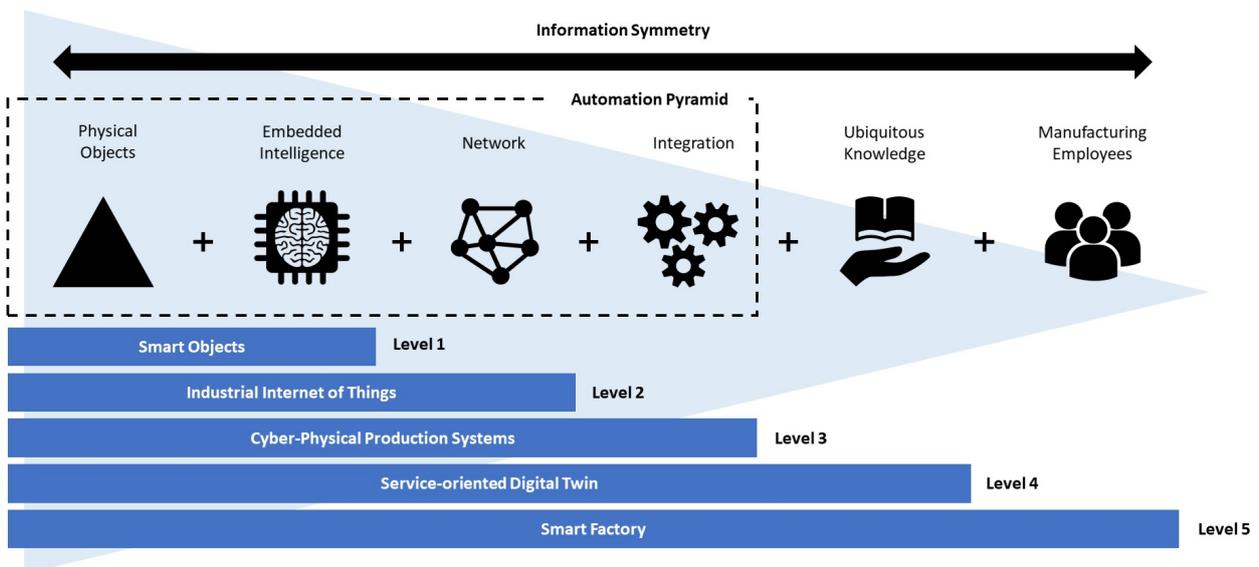

**Figure 1. The Industrial Internet pyramid**

## 3. Approach

Since several high-level service frameworks have been recently proposed (Li et al., 2017) but no clear implementation or application studies have been developed yet, appropriate models and guidelines for implementing the Industrial Internet pyramid and developing a Service-oriented Digital Twin are provided in this section. This section proposes:

- a service-oriented model for the Industrial Internet as a methodological framework for a ubiquitous knowledge about the manufacturing system;
- a technological architecture of the Service-oriented Digital Twin and all its components, with a specific focus on a flexible ontology-oriented knowledge structure used for the knowledge representation and on augmented reality technologies combined to a vocal interaction system for an intuitive knowledge retrieval and fruition.

### 3.1. A service-oriented model for the Industrial Internet

Through data acquisition and analysis, physical manufacturing resources can be virtualized (i.e. in their Digital Twin) and their status and behavior can be dynamically tracked, monitored and adjusted by the manufacturing employees (both enterprise managers and shop-floor operators) thanks to a set of apps and services that can be accessed at any place and any time via cloud. A service-oriented model to implement the Industrial Internet pyramid is proposed in Figure 2.

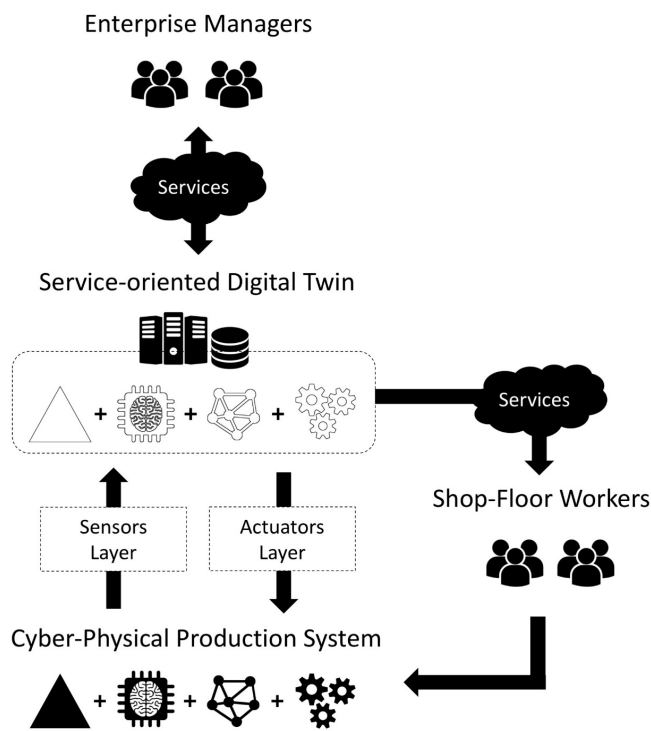

**Figure 2. An Industrial Internet service-oriented model**

Data are real-time collected by the sensors layer for every physical manufacturing resource (e.g. material, machine tools, machining centers, robots, products, software resources, etc.), including basic attributes, real-time status, process parameters, processing progress, maintenance records, operation data of products, as well as demands data, behaviors data and transaction data of users. Moreover, these elements along with some documents (e.g. robot manual, work plans) that can also be associated to a resource to be indexed by the system and later used by users when needed represent the core of the knowledge about the CPPS. The Service-oriented Digital Twin becomes then the virtual representation of the CPPS, which resides in networked computing systems (e.g. computers, servers) and storage systems (e.g. hard-drives). Knowledge dwells in the storage systems and is encapsulated into different standardized services (provided by the computing systems) which enable manufacturing resources and employees achieving interconnection, communication and, therefore, information symmetry. Such services represent information elements (i.e. models as services, algorithms as services, simulation as a service, optimization as a service etc.) that can be accessed by the manufacturing employees to conduct the manufacturing process and related activities (e.g. maintenance, work plan definition, etc.) in a synergistic manner with the physical manufacturing resources. With the characteristics of interoperability and platform independence, various services can be invoked via cloud to request a comprehensive knowledge about the CPPS and use that knowledge to act upon the system.

The Service-oriented Digital Twin offers services spanning over the whole decision-making and control process. Decision-making would greatly benefit of knowledge-empowered managers, who won't monitor directly the CPPS or act directly upon it, but will monitor and interact with the real-time updated Digital Twin of the CPPS. The present model includes four services for the manufacturing managers:

- the *Fault Diagnosis and State Monitoring Service* detects the real-time status, the history and scheduled operations as well as potential sources of ongoing faults of the manufacturing resources based on the sensed and processed data, thus providing managers with a valuable tool to understand and analyze the CPPS's state and health condition.

- the *Prognostics and Scenario Optimization Service* predicts the CPPS's behavior and forecasts the future performance of a possible production scenario according to its real-time state, current and forecasting production data, simulation models, rules, optimization algorithms etc. and user-defined parameters, thus providing a way to verify production capability and capacity, to minimize risks of downtime and costly failure, to identify priorities for maintenance and to extend lifecycles of critical assets.

- the *Manufacturing Scenario Execution Service* converts a possible production scenario into low-level parameters for the CPPS and instructions for the actuators layer, thus providing the manager with a way to implement concretely and swiftly new work plans, new maintenance plans or new production schedules.

- the *Notification Service* alerts the manufacturing manager if an unexpected status or behavior of the CPPS is detected in order to allow them to make timely decisions before faults occur. This service is a proactive version of the Fault Diagnosis and State Monitoring Service, in which the Service-oriented Digital Twin itself notifies the manager of an unpredicted/unsafe behavior.

In association with the manager, the shop-floor operators work alongside the manufacturing system to supervise the advancement of the production and resources' state, to control the manufacturing resources' operating parameters, to carry out activities such as machine setup or maintenance operations. Since employees cannot act remotely upon the systems and resources for safety reasons, the role of the shop-floor operator is relevant to retrieve knowledge from the Service-oriented Digital Twin and act accordingly upon the CPPS. The three services listed below would provide shop-floor operators with a targeted knowledge in order to achieve relevant benefits in the effectiveness and execution of their tasks:

- like the one for the manufacturing manager, the *In-line Fault Diagnosis and State Monitoring Service* can be used to detect the CPPS's state and health condition but the shop-floor operator can investigate deeper into detail the cause of a fault or unexpected behavior (such as to monitor oscillations of temperature, voltage, vibrations to minimize risk and downtime) and can act upon the manufacturing resources timely and informed.

- the *Augmented Assistance & Tutoring Service* retrieves resource-specific information and augmented content to provide knowledge in an interactive manner. A situation-adaptive multimedia tutoring approach enables the shop-floor operator carrying out informed operations and receiving support during the task execution by a vocal assistant and knowledge navigator, thus minimizing the risk of errors and operating with full knowledge of the system's state.

- like the one for the manufacturing manager, the *Notification Service* alerts the shop-floor operators if an unpredicted or unsafe status/behavior of the CPPS is detected, so that they can get closer to the resource of interest and control/operate on it. While the managers are informed and make high-level decisions about how this alert can be handled, the shop-floor operator can directly act upon the resource.

Data and knowledge encapsulated into these services accessed via cloud will enable the manufacturing employees to augment their working experience, to better judge and understand the status of the system, to carry out their operations more effectively, efficiently and safely, to predict coherently future work plans and operations and to meet the demands of customers in a better way.

### 3.2. A technological architecture of a Service-oriented Digital Twin

The technological solution that best match the service-oriented model above proposed is a distributed system of networked components (e.g. computing systems and devices, storage systems), which also includes smart devices (e.g. smartphones, tablets, smartwatches) referred to as remote terminal units (RTUs). RTUs will serve as gateways to the knowledge as they provide the manufacturing employees access to the services permitting cross-layer integration among the CPPS, the Service-oriented Digital Twin and the employees. All the systems in the network coordinate themselves and their actions through the exchange of messages encapsulated into RESTful web services (WS). REpresentational State Transfer (REST) architectural style has been applied because of its greater popularity (Khan et al., 2017) and because it exploits the web-based nature of the system and uses ubiquitous Internet standards (Iarovyi et al., 2016). Moreover, the application of RESTful WS provides easy integration with the tools to manage the CPPS (e.g. a third party ERP system or MES). As the Service-oriented Digital Twin is a distributed network of computing systems providing or consuming web services, a WS mapping has been implemented by using the Constrained application protocol (CoAP), widely used in the Internet of Things domain, into an Enterprise Service Bus implemented into a central server. Its duty is to receive the service request, to route it to the correct producer (the system providing the service) and, once the result is ready, to send it back to the consumer (the RTU requesting the service). The server makes resources (the components of the architecture) available under a URL, and clients access these resources using methods such as GET, PUT, POST, and DELETE. From a developer point of view, CoAP feels very much like HTTP. If the manufacturing employees intend to obtain a value from a sensor in the CPPS via the mobile application, the procedure is not much different from obtaining a value from a Web API. A resource is requested by using an URL like www.digital-twin/api/getMachine/000X/getStatus,

where 000X is the identification number of the machine we want to analyze, which will return a JSON file containing all the information about the machine status.

The technological architecture of the Service-oriented Digital Twin application here proposed is illustrated in **Error! Reference source not found.**. It consists of four main components plus an additional component representing a generic third party system (e.g. specific manufacturing control modules or ERP system) that can be used to provide value adding services to the manufacturing employees (e.g. interaction with the CPPS) or to the enterprise's stakeholders (e.g. delivery time of the 'mass-customized' product based on the real-time status of the plant, use of ERP data for testing with the Digital Twin).

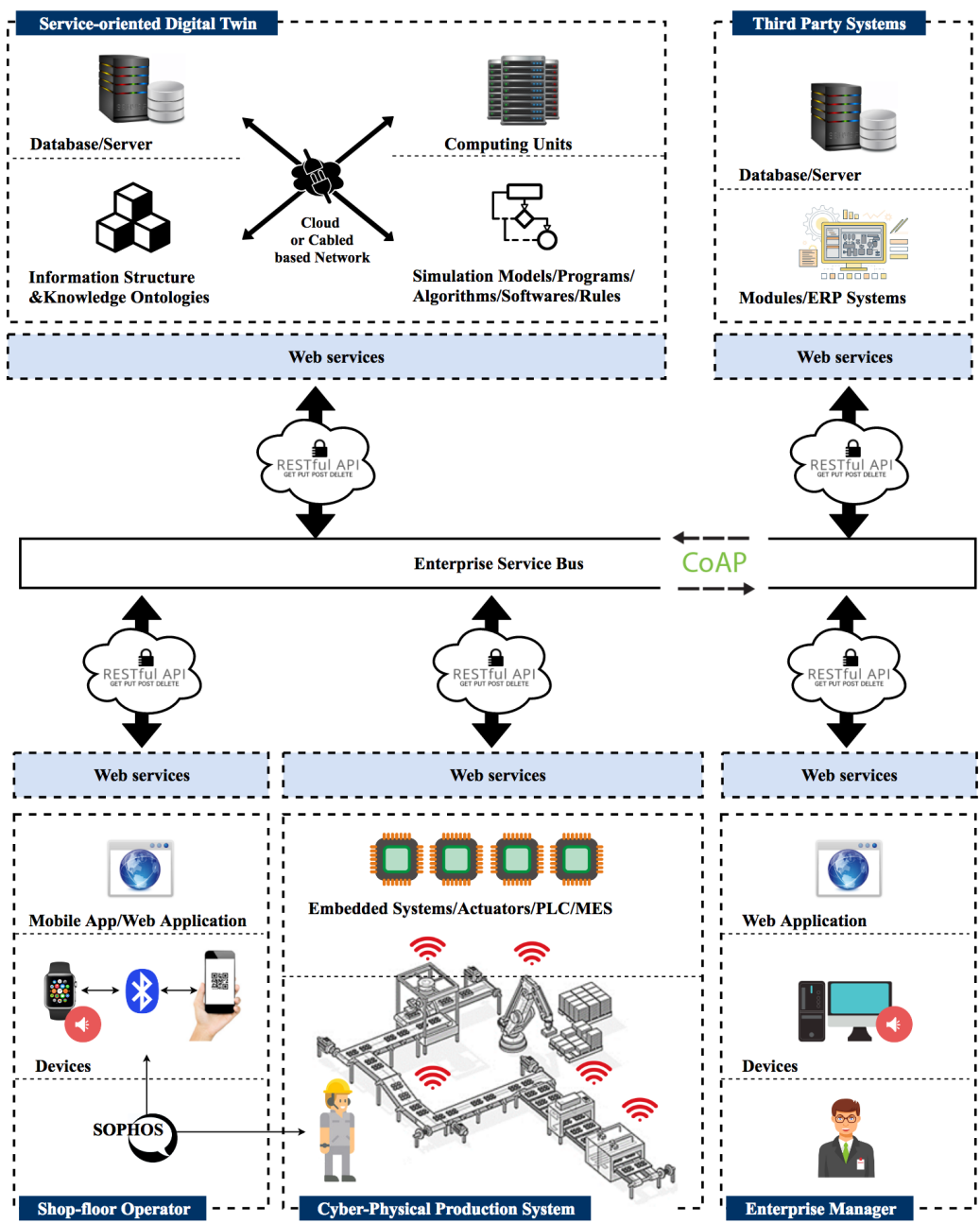

**Figure 3. A technological architecture of a Service-oriented Digital Twin application**

The central element of the architecture is the CPPS, intended as the set of physical manufacturing resources (e.g. material, machine tools, machining centers, robots, products, software resources, etc.) equipped with a local intelligence and connected to a layer of embedded systems, actuators, PLCs and to the Manufacturing Execution System. The CPPS sends seamlessly sensed data to the Service-oriented Digital Twin, which is the real-time synchronized digital reflection of the CPPS. It is (i) self-evolving because its data are updated in real-time (at every time step); (ii) accurate as its behavior reproduces exactly the one of the CPPS thanks to high-fidelity functional models, simulation models, physical models, programs and optimization algorithms, and (iii) knowledge-oriented as different WSs offered by the Service-oriented Digital Twin leverage on the knowledge representation (information structure and knowledge ontologies contained in storage systems) and knowledge creation (through different agent-based and discrete-event simulation models, programs, heuristics-based optimization algorithms). RTUs serving here as gateways to the knowledge are represented by a web application or mobile application installed on the mobile device (e.g. smartphone and smartwatch) of the shop-floor operator and by a web application that can be used by the manufacturing manager to connect through a computer.

### 3.2.1. A flexible ontology-based knowledge structure

The need of a flexible knowledge structure reusable for different manufacturing processes and CPPS configurations as well as resilient to changes in the CPPS over time is a relevant issue that can undermine the functionality of the Service-oriented Digital Twin and has been addressed in this architecture. To map all the raw and processed data generated by the CPPS or by the Service-oriented Digital Twin, a flexible and configurable flexible ontology-based knowledge structure has been designed, so that no programming expertise would be needed to add new data stream to the knowledge structure or modify the knowledge structure according to the user's need. An extracted section of the ontology-based knowledge structure is provided in Figure 4. In this ontology, the 'Item' class represents the generic industrial asset or manufacturing resource whose status at a specific time is defined as a combination of:

- a predefined set of static attributes that do not change over time (unless they are modified by the system's administrator), such as unique ID, name, description, category, the collection of previous state objects, multimedia files such as photos, videos, audio or generic files (e.g. PDF), 3D models, further info, the last update time;

- a variable number of dynamic custom attributes that can be defined by using a GUI depending on the type of asset and linked to data streams coming from sensors in the CPPS. Examples may be a 'cut angle' for a band saw machine or the 'speed' for a lathe, or 'oscillation' and 'operating temperature' for an electric engine.

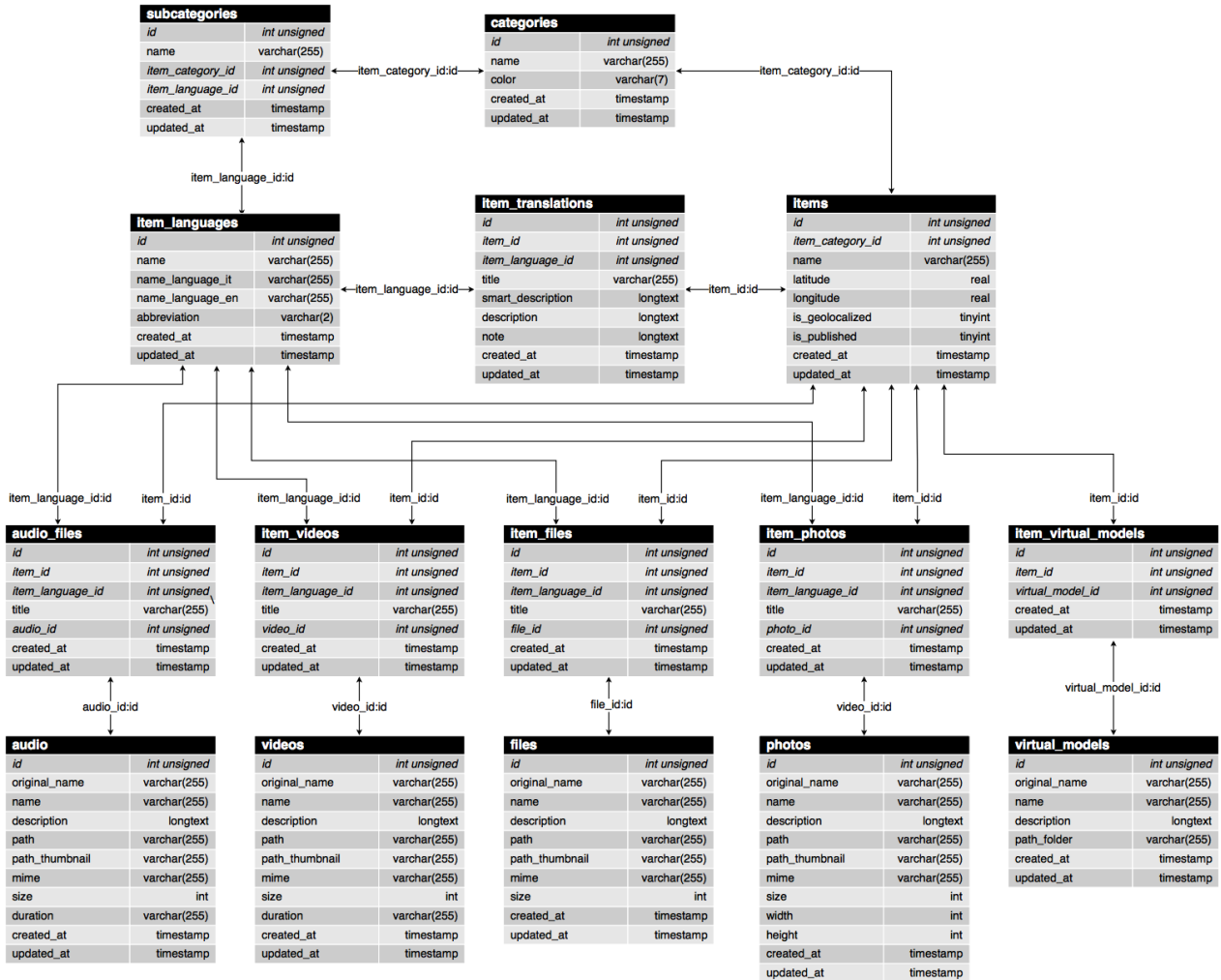

**Figure 4. An extracted section of the ontology-based knowledge structure**

To this end, the Laravel open-source PHP web framework was integrated in the web server to provide the APIs for a web application back-end through which the system administrator will be able to set up the ontology-based knowledge structure. A GUI has been developed (some screenshots are provided in Figure 5) to allow the system administrator creating the Digital Twin of the resources, adding information about the real items and connect the Digital Twin to the real physical manufacturing resource by using a QR code. This can be generated directly from this interface thanks to the embedded jQuery.qrcode plug-in so that it can be printed and attached to the real machine. The QR code is therefore the bridge between the RTU and the Service-oriented Digital Twin. By using the configuration panel for each manufacturing resource, custom attributes

(even depending on the type of sensed data) can be added/edited/removed: primitive type variables (e.g. double, integer, long, string), textual information, photos, videos, audio or generic files, and 3D models. In case of a primitive type attribute, it can be linked to a data stream coming from the CPPS through the Enterprise Service Bus.

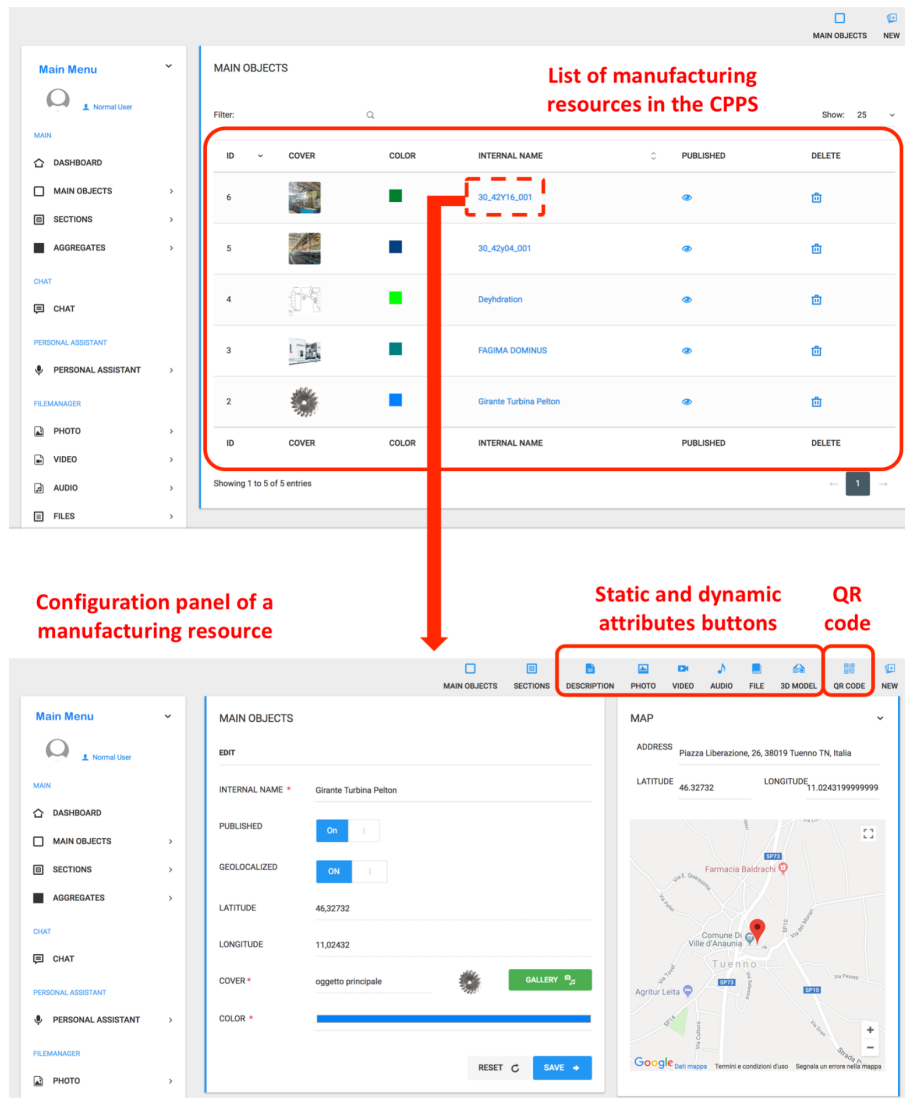

**Figure 5. Configuring the Service-oriented Digital Twin: the system administrator's system**

### 3.2.2. A technological solution for an advanced ubiquitous knowledge fruition

The application prototype for the shop-floor operator has been designed to provide information in a fast and efficient way by leveraging on the technological structure depicted in Figure 6. It has been developed as an Android application able to recognize the QR code and to request a service by clicking on the screen or by vocal interaction. The request will elicit a response in JSON format that will be interpreted by the smart device and converted in usable format (e.g. text, images, audio, video, 3d models). Moreover, the server is equipped with Apache Solr (a java-based open-source platform for full text based search, hit highlighting, faceted search, dynamic grouping database integration and rich documents - doc, docx, pdf - management) and the plug-in Lucene (an open-source high-performance and java-based search engine that provides APIs to retrieve information based on inverted index based algorithms) which enables them to provide an intelligent '4.0' knowledge navigation and vocal interaction service (as already proposed and tested in Longo et al., 2017). Since Solr and Lucene are based on "key-value" non-relational databases, searches are much more efficient and fast compared to traditional object-relational mapping. This design solution is based on the Android's Speech-To-Text (STT) and Text-To-Speech (TTS) features that are solid and robust systems and have been proved to provide a 2 seconds average delayed response and a 95% reliability, i.e. the system will be able to answer correctly simple questions by using keyword recognition and complex questions by a sentence semantic analysis.

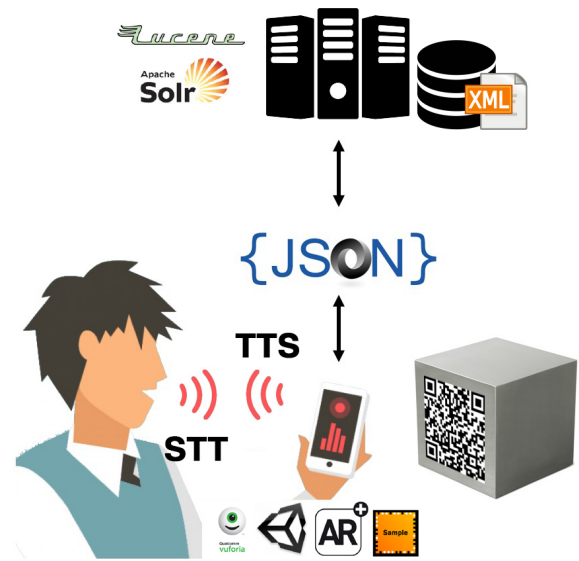

**Figure 6. A technological structure for an advanced ubiquitous knowledge fruition**

The shop-floor operator can use this application to access from the RTU (e.g. a smartphone) the knowledge about the CPPS. This knowledge can include for example the following information (from left to right in Figure 7):

- general information about the machine;
- a 3D virtual exploration of the constructive details of the machine (Augmented Assistance & Tutoring Service);
- which is the most worn element of the machine to be replaced in the next days, in this case the safety switch (In-line Fault Diagnosis and State Monitoring Service);
- a video about the operations of a machine element from the last manufacturing process (Augmented Assistance & Tutoring Service);

- a list of the most critical operations carried out by the machine in the past (In-line Fault Diagnosis and State Monitoring Service).

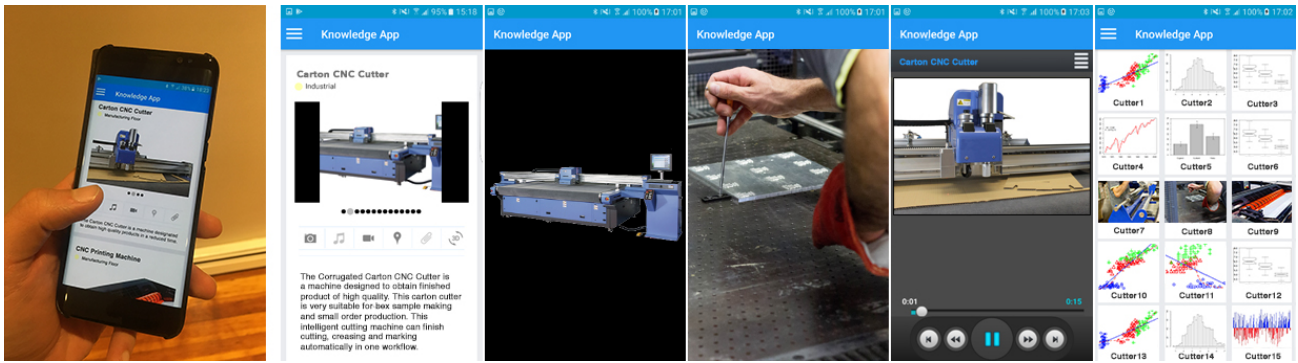

**Figure 7. Advanced knowledge access and fruition by the shop-floor operator**

## 4. Application studies and discussion of the results

An application prototype has been developed and deployed in two industrial manufacturing case studies (a large and a small enterprise) to show whether the hypothesis of this work - a Ubiquitous Knowledge through apps and services provided by a Service-oriented Digital Twin can enhance the performance of Industry 4.0 initiatives in terms of time, costs and quality - is demonstrated by statistically significant quantitative results.

The implementation of the present solution is not instantaneous but required a strong collaborative effort between the research group and the companies involved. After strong efforts in structuring the available knowledge (further details are provided in the next paragraphs), the proper hardware has been installed on site. A central server has been used, where Apache HTTP Server and Apache SOLR have been installed together with a MySQL database. The workstation for the enterprise manager and the Android-based smartphone for the operator on site have been also provided to the companies and connected to the enterprise's intranet.

### 4.1. The case of a large enterprise: Baker Hughes General Electric

The application study for a large enterprise has been conducted in the Baker Hughes (a General Electric company) plant (BHGE) located in Vibo Valentia (Italy), global excellence center in the production of equipment and products for the Oil & Gas industry. GE is world-wide considered a pioneer in the implementation of Industry 4.0 practices and technologies, therefore it can be considered in toto an Industry 4.0-ready company and a perfect test-bed for our hypothesis. In the perspective of a continuous improvement, the driving force for BHGE to be a test-bed in this study was the willingness to exploit the concept of a Ubiquitous Knowledge to enhance the preventive and corrective maintenance activities related to two machines, a fin tube machine (see Figure 8.a) and a milling machine (see Figure 8.b), which are part of the production line of finned tubes (depicted in Figure 8.c).

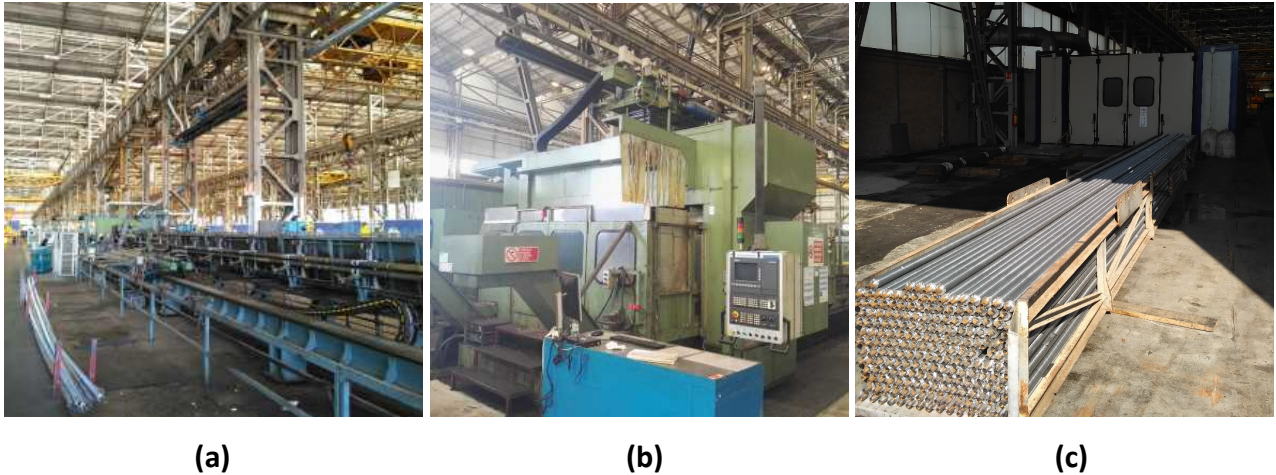

<div align="center">(a)          (b)          (c)</div>

**Figure 8. The fine tube machine (a), the milling machine (b) and the final finned tubes (c)**

One of the biggest challenges for large enterprises is structuring, managing and making available to the employees large volumes of information across different organizational levels. This step is extremely facilitated by the developed ontology-based knowledge structure, which has been proved to be in the implementation stage extremely powerful in creating a faithful Digital Twin of the manufacturing items. The only effort in this phase was devoted to collect all the data, information and multimedia content related to the two machines. Manuals, 3D construction models, security norms, standard maintenance plans and mode of use have been entered in different formats (e.g. single data, excel files, word or PDF files, images, videos) into the Digital Twin of the manufacturing resource through the administrator's views and panels. Custom parameters related to the machines' maintenance operations and considered crucial by the shop-floor operators maintenance operators (e.g. operating temperature, oscillations and tools thickness) have been also created in the Digital Twin of the physical resources and connected to the data streams coming from sensors placed on the machine itself. These parameters will allow them to monitor real-time the machines' health status and therefore to plan more efficiently and automatically the Maintenance Work Plans (MWPs) of the two machines as function of the Production Schedules.

MWPs are generated by matching "manually" the customers' orders, the production schedules and the standard maintenance suggested in the machines' manuals. The current approach is too time-consuming because of the difficulty to retrieve the necessary information. Maintenance activities in large enterprises are usually managed and executed by third-party companies (as in the case of BHGE) that do not know the production schedules or customers' orders, thus causing delays and inefficiencies in the manufacturing processes. For example, if a MWP contains 70 operations and for this reason a machine should remain idling for 5 days, then that MWP may result unfeasible because a high production level is expected in that period.

The use of the Service-oriented Digital Twin services by the manufacturing managers and the maintenance operators is expected to enable them to generate a feasible MWP quickly, thus keeping high performance levels into the manufacturing system. While the Fault Diagnosis and State Monitoring Service can be used to collect real-time data about the current and predicted health status of the machines, the Prognostics & Scenario Optimization Service can be used to carry out an automated generation of the MWP based on the data from the ERP (e.g. customers' orders, production schedules) and the Service-oriented Digital Twin (e.g. maintenance data). The description of the optimization approach for generating the MWP is not actually the focus of this work as this application first aims at assessing the benefits deriving from a quick and valuable knowledge access and fruition compared to the current information asymmetry between the enterprise manufacturing managers and the external maintenance operators.

Table 1 reports the main summary statistics of the estimated time needed to revise the MWP of the two machines before and after the deployment of the Service-oriented Digital Twin based application prototype (for convenience in the discussion, we refer to Group 1 and Group 2 respectively) over 50 observations in both cases. We first determined whether the samples are normally distributed. The Anderson-Darling Normality test shows that data can be assumed normally distributed only in the case of Group 2 for the fin tube machine (p-value is less than 0.05) but not in the other cases, therefore we assume the data non normal distributed. A 2 variances one-sided test has been conducted to check whether the variances of the two groups per each machine differ as the sample data have been collected randomly and independently, data are not severely skewed (as shown in Table 1) and the sample size is appropriate ($N = 50$). The null hypothesis is:

$$H_0: \sigma_1{}^2/\sigma_2{}^2 = k \,, k = 1$$

meaning that the ratio between the first population variance ($\sigma_1{}^2$) and the second population variance ($\sigma_2{}^2$) is equal to the hypothesized ratio ($k = 1$), while the alternative hypothesis is:

$$H_1: \sigma_1{}^2/\sigma_2{}^2 > k$$

For non-normal distributed data, the Bonett's method or Levene's method with a significance level $\alpha = 0.05$ have proved that the two groups for the two machines have different variances ($p-value \cong 0$). A One-Way Analysis of Variance and a Games-Howell non-parametric post-hoc test has been eventually performed to check whether the means of the two groups are different (the null hypothesis is $\mu_1 = \mu_2$). Even if the samples are not both from a normal distribution, the sample size meet the requirements for a reliable use of the one-way ANOVA. The p-value resulted to be almost null in both cases, thus strengthening the fact that the null hypothesis should be rejected and that the two means are significantly different (and therefore the mean in Group 2 is significantly lower than the mean in Group 1) for both machines.

**Table 1. Statistical analysis on the Maintenance Work Plan generation time**

| | **Fin Tube Machine** | | **Milling Machine** | |
| --- | --- | --- | --- | --- |
| | **Group 1** | **Group 2** | **Group 1** | **Group 2** |
| Summary Statistics | | | | |
| N | 50 | 50 | 50 | 50 |
| Mean | 83.34 | 6.25 | 49.36 | 4.49 |
| Standard Deviation | 36.25 | 4.02 | 14.01 | 2.59 |
| 95% CI | (73.04; 93.64) | (5.11; 7.39) | (45.38; 53.34) | (3.75; 5.25) |
| Minimum | 21.12 | 0.42 | 21.04 | 0.25 |
| Q1 | 56.81 | 2.46 | 39.54 | 2.62 |
| Median | 81.58 | 6.46 | 47.97 | 3.97 |
| Q3 | 107.83 | 9.25 | 57.31 | 6.40 |
| Maximum | 154.53 | 15.17 | 94.24 | 11.21 |
| Skewness | 0.23 | 0.39 | 0.67 | 0.39 |
| Anderson-Darling Normality Test | | | | |
| A-Squared | 0.51 | 0.83 | 0.35 | 0.42 |
| p-value | 0.193 | 0.031 | 0.463 | 0.310 |
| 2 Variances One-sided Test | | | | |
| Test statistics (Bonett) | 149.84 | | 43.12 | |
| p-value (Bonett) | 0.000 | | 0.000 | |
| Test statistics (Levene) | 87.02 | | 48.51 | |
| p-value (Levene) | 0.000 | | 0.000 | |
| One-Way ANOVA and Games-Howell Pairwise Comparison | | | | |
| Difference of Means | -77.09 | | -44.87 | |
| 95% CI | (-87.45; -66.74) | | (-48.91; -40.83) | |
| T-value | -14.95 | | -22.27 | |
| Adjusted p-value | 0.000 | | 0.000 | |

The results clearly show that the generation of the MWP by the Service-oriented Digital Twin and the immediate access to this new knowledge by the enterprise managers and maintenance operators (Group 2) allows the company to achieve greater improvements in time compared to the current approach (Group 1). Considering that a revision and re-generation of the MWP is needed every time a new order arrives because the production schedule must be modified accordingly, this situation is really frequent and inefficiencies can be relevant even in terms of costs. Since customers' orders and production plans vary constantly, it is hard to predict the benefits that can be obtained over time. If we estimate the cost of an hour of inefficient production per single machine to be 50 €/hour and consider an average number of orders per year for the enterprise under consideration equal to 50, then the cost for inefficient MWP per single order is 101.64 €/order, that means 5 081.77 €/year, which is a considerable economic saving that can be achieved just by using an automated generation of the MWP enabled by the Service-oriented Digital Twin and a full information sharing and access between enterprise manager and external maintenance operators. Moreover, once the MWP has been generated and activated, preventive and corrective maintenance activities may also benefit from a Ubiquitous Knowledge even in the estimated time to execute maintenance operations. Although maintenance activities are executed by a third-party company, reductions of the time to execute such operations do not only provide benefit to the productivity of the maintenance company itself, but also to BHGE, which would benefit of a reduced downtime due to maintenance and reduced maintenance costs.

To assess the impact of a Ubiquitous Knowledge on the maintenance execution time, data on preventive and corrective maintenance operations have been collected before and after the use of the proposed application prototype. The research team has conducted the experimentation on the real machines according to the following design:

- the preventive maintenance operations have been recreated to evaluate the time needed to perform them after the deployment of the application prototype. Historical data have been instead used as an estimation of the time before the use of the application;

- the corrective maintenance operations have been recreated by injecting artificial alarms and asking the maintenance operators to perform the appropriate procedure without and, later, with the support of the Service-oriented Digital Twin. A selection of mechanical, electric and hydraulic/pneumatic alarms for both machines has been used for the purpose of this study and the time needed to fix them has been observed. 60 mechanical alarms, 26 electrical alarms and 40 pneumatic/hydraulic alarms have been injected for the fin tube machine, whereas 90 mechanical alarms, 35 electrical alarms and 5 pneumatic/hydraulic alarms have been recreated for the milling machine.

A summary of the results is provided in Table 2 for the fin tube machine and in Table 3 for the milling machine, in which the cost of the maintenance operator per hour is 25 €/h, that is 0.42 €/min.

In the case of the fine tube machine, it can be observed that the estimated total cost for the preventive maintenance is 14 012.08 €/year, due to 33 629 minutes of operations, whereas a total of 12 462 minutes of machine breakdowns and 5 192.50 €/year are expected for the corrective maintenance activities. In the case of using the proposed application prototype, the operations for preventive maintenance have undergone a strong time reduction, which led to a total amount of about 28 640 minutes and 11 933.33 €/year. Considering the same scenario (same number of alarms), corrective maintenance operations have been subjected to an interesting reduction of the time needed to fix the issue causing the alarm, with a final cost for corrective maintenance equal to 4 013.33 €/year. A 14.84% reduction of the cost is therefore registered for the preventive maintenance of the fin tube machine, while a 22.71% is obtained for the corrective maintenance, with a total saving of 16.96%. In the case of the milling machine, the estimated preventive maintenance cost before the deployment of the application prototype was 8 751.25 €/year, while after its implementation was reduced to 7 074.17 €/year, with a percentage reduction of 19.16%. The corrective maintenance has also undergone a strong improvement with a 21.24% reduction of the time needed to fix the alarms, which resulted into a reduction of the costs for corrective maintenance from 2 452.08 €/year to 1 931.25 €/year. The total economic saving on the maintenance costs of the milling machine is 19.62%, from 11 203.33 €/year to 9 005.42 €/year.

The cost here considered only takes into consideration the cost of the personnel that it is called to repair or maintain the machine (the maintenance operators). Further improvements can be actually obtained because the use of a vocal assistant (Augmented Assistance & Tutoring Service) throughout the maintenance operations can support the operator to carry out correct operations, thus avoiding waste of spare parts or tutoring on the correct use of tools used during the maintenance operations. The significance of the potential improvements on maintenance activities here discussed is even greater if we consider that this economic savings refer only to two machines and not to all the machines, equipment and machining centers of the large enterprise. In addition, BHGE can benefit of a better matching between the MWPs with the production schedules, which can enable them to provide mass customization services to the customers even in case of big orders and commissions. Since such orders are indeed completed after different months, the impact of maintenance operations on the delivery time and customer value added is significant, therefore a better planning of the maintenance operations according to the customers' orders and production schedules will enable the large enterprise to provide greater reliability and better services even to the customers.

**Table 2. Preventive and corrective maintenance on the fin tube machine**

| | | Preventive Maintenance | | | | | Corrective Maintenance | | | Total |
|---|---|---|---|---|---|---|---|---|---|---|
| | **MWP** | **Annual** | **Trimestral** | **Monthly** | **Weekly** | **Total** | **# Alarms** | **Average Time** | **Total** | |
| **Before** | Mechanical [min] | 221 | 1047 | 1076 | 73 | 21117 | 60 | 95 | 5700 | |
| | Electrical [min] | | 473 | | | 1892 | 26 | 57 | 1482 | |
| | Pneumatic/Hydraulic [min] | | 185 | | 190 | 10620 | 40 | 132 | 5280 | |
| | Total [min/year] | 221 | 6820 | 12912 | 13676 | 33629 | | | 12462 | 46091 |
| | Cost [€/year] | 92,08 | 2841,67 | 5380,00 | 5698,33 | 14012,08 | | | **5192,50** | 19204,58 |
| | **MWP** | **Annual** | **Trimestral** | **Monthly** | **Weekly** | **Total** | **# Alarms** | **Average Time** | **Total** | |
| **After** | Mechanical [min] | 184 | 897 | 916 | 59 | **17832** | 60 | 71 | 4260 | |
| | Electrical [min] | | 368 | | | **1472** | 26 | 42 | 1092 | |
| | Pneumatic/Hydraulic [min] | | 163 | | 167 | **9336** | 40 | 107 | 4280 | |
| | Total [min/year] | 184 | 5712 | 10992 | 11752 | 28640 | | | 9632 | 38272 |
| | Cost [€/year] | 76,67 | 2380,00 | 4580,00 | 4896,67 | **11933,33** | | | **4013,33** | 15946,67 |

**Table 3. Preventive and corrective maintenance on the milling machine**

| | | Preventive Maintenance | | | | | Corrective Maintenance | | | Total |
|---|---|---|---|---|---|---|---|---|---|---|
| | **MWP** | **Annual** | **Trimestral** | **Monthly** | **Weekly** | **Total** | **# Alarms** | **Average Time** | **Total** | |
| **Before** | Mechanical [min] | 240 | 1003 | 689 | 100 | 17720 | 90 | 45 | 4050 | |
| | Electrical [min] | 913 | | 105 | | 2173 | 35 | 36 | 1260 | |
| | Pneumatic/Hydraulic [min] | 30 | | 90 | | 1110 | 5 | 115 | 575 | |
| | Total [min/year] | 1183 | 4012 | 10608 | 5200 | 21003 | | | 5885 | 26888 |
| | Cost [€/year] | 492,92 | 1671,67 | 4420,00 | 2166,67 | 8751,25 | | | **2452,08** | 11203,33 |
| | **MWP** | **Annual** | **Trimestral** | **Monthly** | **Weekly** | **Total** | **# Alarms** | **Average Time** | **Total** | |
| **After** | Mechanical [min] | 200 | 823 | 580 | 74 | 14300 | 90 | 35 | 3150 | |
| | Electrical [min] | 770 | | 85 | | 1790 | 35 | 30 | 1050 | |
| | Pneumatic/Hydraulic [min] | 24 | | 72 | | 888 | 5 | 87 | 435 | |
| | Total [min/year] | 994 | 3292 | 8844 | 3848 | 16978 | | | 4635 | 21613 |
| | Cost [€/year] | 414,17 | 1371,67 | 3685,00 | 1603,33 | **7074,17** | | | **1931,25** | **9005,42** |

**4.2. The case of a small and local enterprise producing carton packaging boxes**

Alongside the application in BHGE, another application study has been conducted in a small enterprise producing carton packaging boxes with different dimensions and shapes. The company is considering the possibility to expand horizontally in the packaging for agri-food and cosmetic products and they feel that the new Industry 4.0 technologies and the concept of Smart Factory may help them to produce custom output at a low unit costs and with great production flexibility but they doubt the convenience of the investment. In their production process, the only raw materials are carton sheets, hailing from paper mills ready for a 4-step production process. The carton sheets are first printed (S1) and then cut in the desired shape and dimension (S2). Next, two lateral grips are obtained on a punch cutting machine (S3) and, ultimately, the sheets folded up and glued to form a box (S4). The production of every batch is automated and anticipated by machine set-up operations that introduce delays and frequent errors (reflected in the medium-high production waste rate): activities include, among others, machine setup, quality control checks, tools cleaning, replacing, and material positioning. The objective is to reduce significantly the setup times and the cycle times of each production step, while reducing simultaneously the production waste rate.

Empirical data about setup and cycle times and number of waste products have been collected over a 100 batches production time for 10 different operators in two cases, before and after the deployment of the application prototype installed on a smartphone to enable operators accessing knowledge more intuitively, especially during the setup process. A synthetic view of the average reduction for cycle time and setup time over a batch of products produced by the 10 operators is given in Figure 9, while the average batch cycle times and setup times before and after the use of the application prototype for the 10 operators is depicted in Figure 10. The figure highlights that (as expected) operators using the application have, on average, better performances and, as a result, the setup time and cycle time decrease. In particular, a 7,57% decrease of the cycle time (from 240,4 min to 223,1 min) has been obtained thanks to a more consistent 28,62% decrease of the setup time, which dropped from 63,8 min to 46,6 min. The most relevant improvement has been obtained in the first production step, where the setup time decreased on the average of 44,5% (from 18,5 min to 10,3 min). This has been generally due to the fact that in this very first production step, the Tutoring Service and Augmented Assistance provided by the application enabled the operator to speed up the operations, to be more precise and coherent with the procedure without delays due to lacking of knowledge ("how to setup the printing machine for the next batch?") and information asymmetry ("which is the next batch to produce?"). Similarities with the case of the large enterprise in terms of need of no information asymmetry between the shop floor and the customers can be identified. The other production steps have undergone a more modest but substantial improvement mainly thanks to the In-line Diagnostics and Condition Monitoring Service that the operators were used to request during the process to continuously monitor the process parameters and performances. In S2, the setup time has been reduced by 15,6%, which corresponds to a bit less than 2 minutes (from 11,4 to 9,6 min), with a 3,1% reduction of the S2 cycle time. Similarly, the S3 setup time was cut by 17,1% (2,7 min) and the S3 cycle time by 3,4%. A slight greater impact has been registered for S4, which experienced a 24,9% reduction of its setup time (around 4,5 min) and a 10,7% reduction of its cycle time.

The statistical significance of the results is confirmed by the Fisher's Least Significance Difference (LSD) test, in which we tested the null hypothesis "the means of the batch cycle times and setup times before and after the use of the application prototype come from the same population" for each operator both for the setup time and for the cycle time of each production step as well as for the whole process. According to the LSD method, the pair of means would be declared significantly different if:

$$|\overline{y_1} - \overline{y_{4.1}}| > LSD$$

where $\overline{y_1}$ and $\overline{y_2}$ are the mean values of the batch cycle times and setup times before and after the use of the application prototype while the LSD is calculated as follows

$$LSD = t_{\frac{\alpha}{2}, N-a} \sqrt{\frac{2MS_E}{n}}$$

where $t$ is the t-distribution, $MS_E$ is the Mean Square Error, 0.025 is the half of the confidence level, 98 is given by the number of observations 100 minus 2 (the number of compared scenarios - before and after the deployment of the application).

Results reported in Table 4 show that the LSD is always smaller than the difference (in absolute value) between the means of the setup times (or cycle times), indicated by δ, thus corroborating our hypothesis that the observed differences are not due to sampling errors.

A necessary consideration is required for taking into consideration the fact that lower setup times and cycle times for the batch production may result into lower production standards and, consequently, higher production waste. However, this does not occur in this case because the reduction is due to a better knowledge fruition during the production process. The average reduction of the production waste mean rate is given in Figure 11, while the average production waste mean rate before and after the use of the application prototype for the 10 operators is depicted in Figure 12. Even in this case, the reduction of the production waste mean rates is statistically significant as verified by the LSD test reported in Table 5, in which the tested null hypothesis is "Means of the batch production waste rate before and after the use of the application prototype come from the same population".

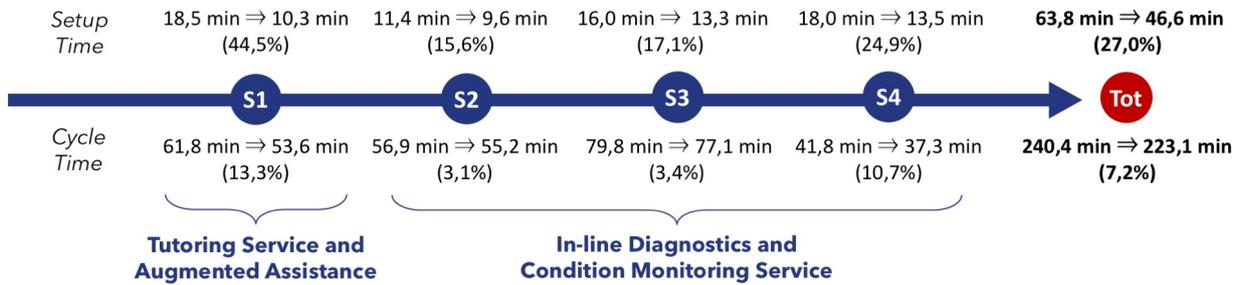

**Figure 9.** Average reduction of the batch cycle times and setup times before and after the use of the application prototype

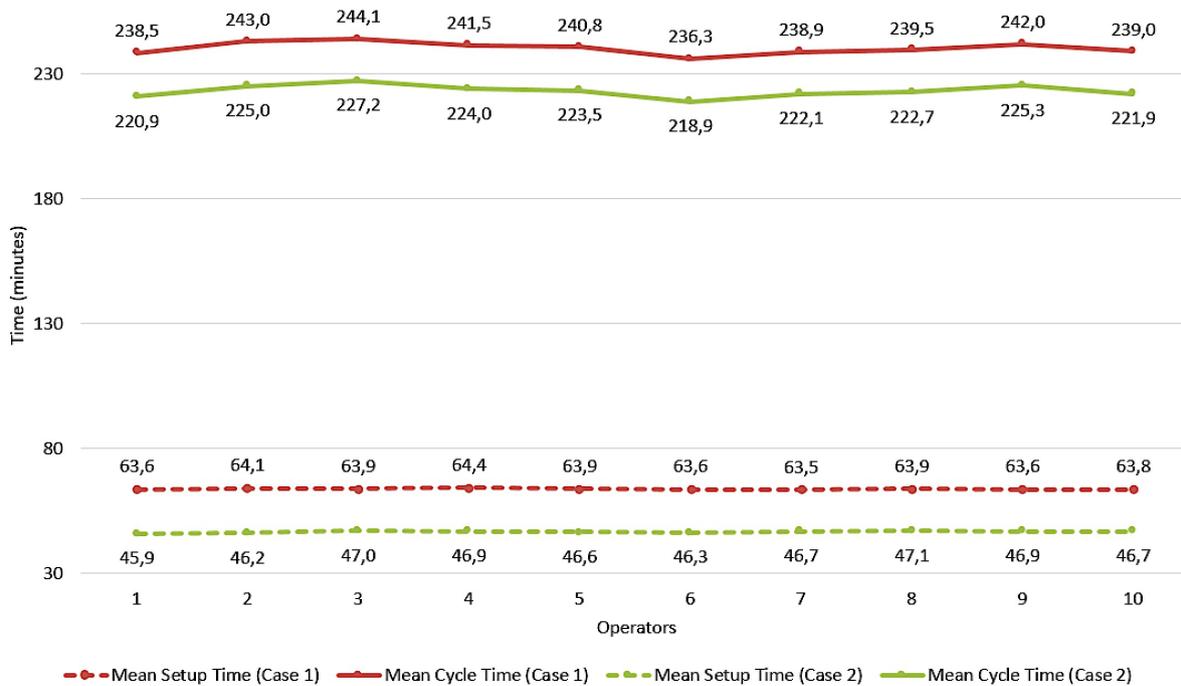

**Figure 10.** Batch cycle times and setup times before and after the use of the application prototype for the 10 operators

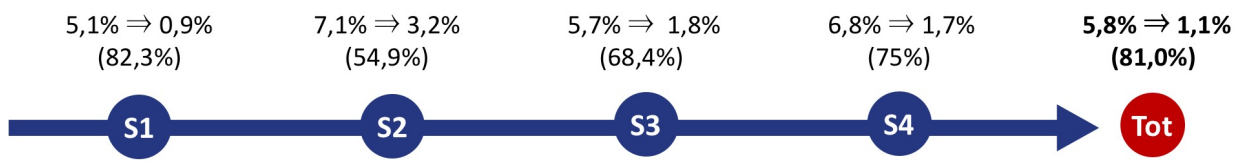

**Figure 11.** Average reduction of the batch production waste rate before and after the use of the application prototype

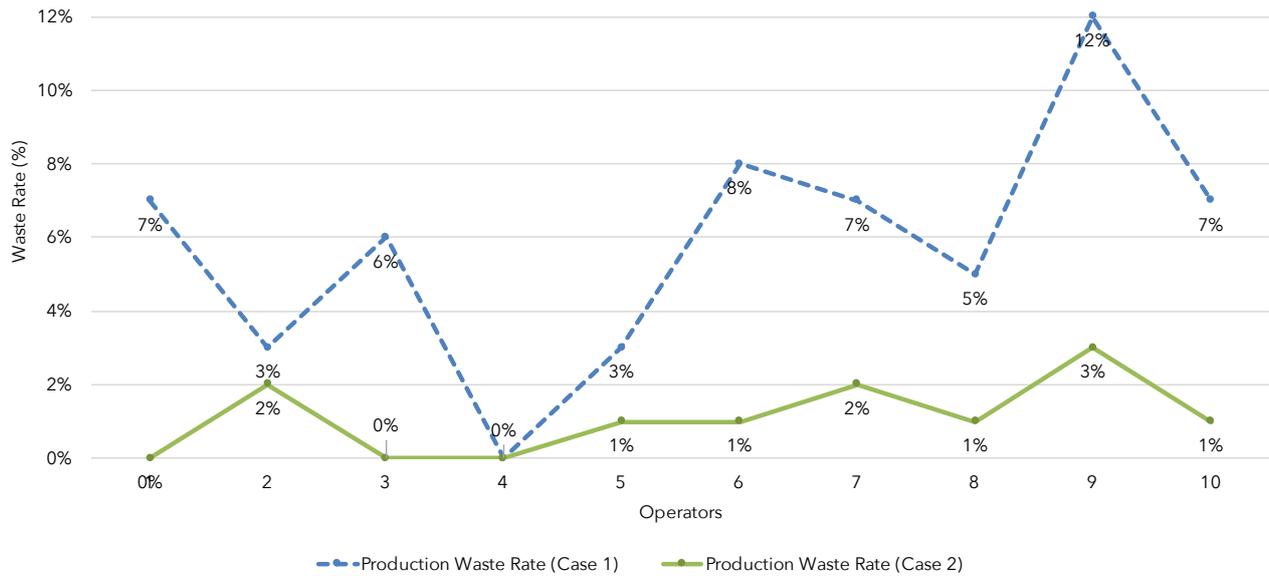

**Figure 12. Batch production waste rate before and after the use of the application prototype for the 10 operators**

**Table 4. Fisher's LSD analysis on cycle times and setup times before and after the use of the application prototype**

| | | S1 | | S2 | | S3 | | S4 | | T | |
|---|---|---|---|---|---|---|---|---|---|---|---|
| | | LSD | δ | LSD | δ | LSD | δ | LSD | δ | LSD | δ |
| **Op1** | **cycle time** | 0,08 | 0,09 | 0,02 | 0,02 | 0,03 | 0,03 | 0,01 | 0,05 | 0,09 | 0,18 |
| | **setup time** | 0,01 | 0,09 | 0,00 | 0,02 | 0,00 | 0,03 | 0,01 | 0,05 | 0,01 | 0,18 |
| **Op2** | **cycle time** | 0,06 | 0,09 | 0,02 | 0,02 | 0,02 | 0,03 | 0,01 | 0,05 | 0,06 | 0,18 |
| | **setup time** | 0,01 | 0,09 | 0,00 | 0,02 | 0,00 | 0,03 | 0,01 | 0,05 | 0,02 | 0,18 |
| **Op3** | **cycle time** | 0,05 | 0,09 | 0,02 | 0,02 | 0,02 | 0,03 | 0,01 | 0,04 | 0,05 | 0,17 |
| | **setup time** | 0,01 | 0,09 | 0,00 | 0,02 | 0,00 | 0,03 | 0,01 | 0,04 | 0,01 | 0,17 |
| **Op4** | **cycle time** | 0,05 | 0,08 | 0,02 | 0,02 | 0,02 | 0,03 | 0,01 | 0,04 | 0,06 | 0,17 |
| | **setup time** | 0,01 | 0,08 | 0,00 | 0,02 | 0,00 | 0,03 | 0,01 | 0,04 | 0,01 | 0,17 |
| **Op5** | **cycle time** | 0,06 | 0,08 | 0,02 | 0,02 | 0,02 | 0,02 | 0,01 | 0,05 | 0,07 | 0,17 |
| | **setup time** | 0,01 | 0,08 | 0,00 | 0,02 | 0,00 | 0,02 | 0,01 | 0,05 | 0,01 | 0,17 |
| **Op6** | **cycle time** | 0,06 | 0,08 | 0,02 | 0,02 | 0,02 | 0,03 | 0,01 | 0,04 | 0,06 | 0,17 |
| | **setup time** | 0,01 | 0,08 | 0,00 | 0,02 | 0,00 | 0,03 | 0,01 | 0,04 | 0,02 | 0,17 |
| **Op7** | **cycle time** | 0,06 | 0,08 | 0,02 | 0,02 | 0,02 | 0,03 | 0,01 | 0,05 | 0,07 | 0,17 |
| | **setup time** | 0,01 | 0,08 | 0,00 | 0,02 | 0,00 | 0,03 | 0,01 | 0,05 | 0,01 | 0,17 |
| **Op8** | **cycle time** | 0,05 | 0,08 | 0,02 | 0,01 | 0,02 | 0,03 | 0,01 | 0,05 | 0,06 | 0,17 |
| | **setup time** | 0,01 | 0,08 | 0,00 | 0,01 | 0,00 | 0,03 | 0,00 | 0,05 | 0,01 | 0,17 |
| **Op9** | **cycle time** | 0,05 | 0,08 | 0,02 | 0,02 | 0,02 | 0,03 | 0,01 | 0,04 | 0,06 | 0,17 |
| | **setup time** | 0,01 | 0,08 | 0,00 | 0,02 | 0,00 | 0,03 | 0,01 | 0,04 | 0,01 | 0,17 |
| **Op10** | **cycle time** | 0,05 | 0,08 | 0,02 | 0,02 | 0,02 | 0,03 | 0,01 | 0,05 | 0,06 | 0,17 |
| | **setup time** | 0,01 | 0,08 | 0,00 | 0,02 | 0,00 | 0,03 | 0,01 | 0,05 | 0,02 | 0,17 |

**Table 5. Fisher's LSD analysis on production waste rate before and after the use of the application prototype**

| | S1 | | S2 | | S3 | | S4 | | T | |
|---|---|---|---|---|---|---|---|---|---|---|
| | $\mu_1$ | $\mu_2$ | $\mu_1$ | $\mu_2$ | $\mu_1$ | $\mu_2$ | $\mu_1$ | $\mu_2$ | $\mu_1$ | $\mu_2$ |
| **Op1** | 0,05 | 0,02 | 0,09 | 0 | 0 | 0,01 | 0,09 | 0,02 | 0,07 | 0 |
| **Op2** | 0,06 | 0 | 0,04 | 0,03 | 0,02 | 0,03 | 0,06 | 0,01 | 0,03 | 0,02 |
| **Op3** | 0,04 | 0 | 0,06 | 0,07 | 0,06 | 0,02 | 0,07 | 0 | 0,06 | 0 |
| **Op4** | 0,04 | 0 | 0,06 | 0,03 | 0,1 | 0,01 | 0,11 | 0 | 0 | 0 |
| **Op5** | 0,05 | 0,04 | 0,1 | 0,02 | 0,08 | 0,02 | 0,04 | 0,02 | 0,03 | 0,01 |
| **Op6** | 0,06 | 0,01 | 0,06 | 0,02 | 0,07 | 0,02 | 0,04 | 0,02 | 0,08 | 0,01 |
| **Op7** | 0,08 | 0 | 0,09 | 0,05 | 0,07 | 0,02 | 0,05 | 0,03 | 0,07 | 0,02 |
| **Op8** | 0,04 | 0 | 0,09 | 0 | 0,05 | 0 | 0,09 | 0,01 | 0,05 | 0,01 |
| **Op9** | 0,05 | 0,01 | 0,06 | 0,06 | 0,08 | 0,04 | 0,07 | 0,02 | 0,12 | 0,03 |
| **Op10** | 0,04 | 0,01 | 0,06 | 0,04 | 0,04 | 0,01 | 0,06 | 0,04 | 0,07 | 0,01 |
| **Media** | 0,051 | 0,009 | 0,071 | 0,032 | 0,057 | 0,018 | 0,068 | 0,017 | 0,058 | 0,011 |
| **LSD** | 0,0199 | | 0,0335 | | 0,0353 | | 0,0286 | | 0,0376 | |
| **δ** | 0,042 | | 0,039 | | 0,039 | | 0,051 | | 0,047 | |

## 5. Conclusions & further research

In the context of a more human-centric Industry 4.0, the Industrial Internet pyramid proposed in this study argues that an enterprise's transformation from an Automated Factory into a Smart Factory will occur when there will be no more information asymmetry within and among technology, process and people, thus achieving a fully (or quasi-fully) coupled human-machine interaction. The complexity of this transformation resides not only in the lack of clear implementation guidelines in literature and of structured information but also in enabling and convincing enterprises and operators (more familiar with legacy systems and traditional procedures) to easily harness this knowledge. Industrial manufacturing environments have indeed demonstrated that they cannot do without the human component, who is now turning from traditional blue-collar workers into knowledge-empowered workers, as envisioned in World Economic Forum (2015). The new perspective proposed in this study aims at demonstrating that a technocratic view of a Smart Factory is no longer sufficient and further improvement of manufacturing systems' performance may be obtained if enterprises climb the Industrial Internet pyramid up to the top and Industry 4.0 initiatives are backed up by a ubiquitous knowledge, which guarantees no information asymmetry between the CPPS and manufacturing employees. The two application studies proved how the potential of current technology-driven Industry 4.0 initiatives have not been fully exploited yet because of a lacking ubiquitous knowledge oriented proneness in manufacturing enterprises and because mere technological change is easily reproducible by the competitors and does not provide any value added or competitive advantage to the company in a worldwide context. This paper aims at making a step forward, leading to an incremental improvement in the vision of the Smart Factory, by proposing the Industrial Internet pyramid and developing the prototype of a Service-oriented Digital Twin that significantly supports the manufacturing employees' daily job in the shop floor of two companies. While the return on investment can be easily calculated in case new technologies, innovative robots or automated smart machines are purchased and deployed in a production line (that is the bottom part of the industrial internet pyramid), enterprises are reluctant in investing money and efforts in the upper part of it (bridging the production system and employees through a seamless bidirectional information stream) because the return on investment is often uncertain. The promising results of this application study show instead how statistically significant improvement of the production and business performance can be achieved, which can be hopefully of inspiration for any type of enterprise. Current research efforts are devoted to extend the capabilities of the intelligent vocal assistant to recognize and answer correctly several other commands or requests. Furthermore, the statistical association between the knowledge management efforts in an enterprise and its performance as well as a prediction of the time needed to have a significant return on investment are under investigation.

In view of this, operating and maintaining industrial assets is just one field of application. The Service-oriented Digital Twin can be the source of several new knowledge-driven business opportunities. This knowledge can be also conveniently provided to the enterprise's stakeholders, such customers, suppliers, partners, etc., to achieve a full organization-wide transformational change or provide innovative services to the customers in the perspective of a mass customization production strategy. In a Smart Factory, production scheduling or maintenance operations can be integrated with sales orders so that the manufacturing employees can real-time plan or reschedule certain activities according to the new production timetable and levels. This aspect is even more crucial when maintenance operations involve outsourcing strategies. It is not unusual the case of maintenance services provided by a third-party company that may be strongly interested in receiving in an intuitive and quick manner information about the possibility to carry out maintenance operations on a machine (or group of machines) based on the expected production levels. An interesting development of the present work could be the application of the Industrial Internet pyramid outside of the manufacturing environment in a more comprehensive way across the enterprise and organizational levels. If we extend the vision of reducing the information asymmetry also among the parties of a supply chain, Service-oriented Digital Twins equipped with intelligent vocal assistants may become the hinges between a supplier and a customer and may streamline and make the supply process more efficient. In this sense, the combination of the Digital Twins with trusted data sharing technologies (such as blockchain) could pave the way to a new wave in supply chain studies. This last aspect of trusted data sharing technologies is extremely relevant in a time when knowledge must be protected by potential cyber attacks. As more industrial smart assets connect to the internet and Service-oriented Digital Twins become popular, cyber attackers may target IT industrial systems more and more often, taking advantage of virtual means to steal organizational knowledge, engage in industrial espionage or extortion or even take control over the physical systems. Therefore, extensive research efforts and practices to ensure proper strong cybersecurity measures and actions to increase knowledge resiliency must be considered.